# Developing the Technique of Measurements of Magnetic Field in the CMS Steel Yoke Elements With Flux-Loops and Hall Probes

V. I. Klyukhin, *Member, IEEE*, D. Campi, B. Curé, A. Gaddi, H. Gerwig, J. P. Grillet, A. Hervé, R. Loveless, and R. P. Smith

*Abstract*—Compact muon solenoid (CMS) is a general-purpose detector designed to run at the highest luminosity at the CERN large hadron collider (LHC). Its distinctive features include a 4 T superconducting solenoid with 6 m diameter by 12.5 m long free bore, enclosed inside a 10000-ton return yoke made of construction steel. Accurate characterization of the magnetic field everywhere in the CMS detector, including the large ferromagnetic parts of the yoke, is required. To measure the field in and around ferromagnetic parts, a set of flux-loops and Hall probe sensors will be installed on several of the steel pieces. Fast discharges of the solenoid during system commissioning tests will be used to induce voltages in the flux-loops that can be integrated to measure the flux in the steel at full excitation of the solenoid. The Hall sensors will give supplementary information on the axial magnetic field and permit estimation of the remanent field in the steel after the fast discharge. An experimental R&D program has been undertaken, using a test flux-loop, two Hall sensors, and sample disks made from the same construction steel used for the CMS magnet yoke. A sample disc, assembled with the test flux-loop and the Hall sensors, was inserted between the pole tips of a dipole electromagnet equipped with a computer-controlled power supply to measure the excitation of the steel from full saturation to zero field. The results of the measurements are presented and discussed.

## I. Introduction

THE COMPACT muon solenoid (CMS) is a general-purpose detector designed to run at the highest luminosity at the CERN large hadron collider (LHC). Its distinctive features include a 4 T superconducting solenoid with 6 m diameter by 12.5 m long free bore, enclosed inside a 10 000-ton yoke made of construction steel. The yoke consists of five dodecagonal three-layered barrel wheels and three end-cap disks at each end, comprised of steel plates up to 630 mm thick which return the flux of the solenoid and serve as the absorber plates of the muon detection system [1], [2].

A three-dimensional (3-D) magnetic field model of the CMS magnet has been prepared [3] for utilization during the engineering phase of the magnet system and early physics studies of the anticipated performance of the detector, as well as for track parameter reconstruction when the detector begins operation. This model uses "averaged" permeability values measured in many samples taken from the steel plates.

It is desirable to provide a direct measurement of the magnetic flux density in selected regions of the yoke to reduce the uncertainty in using the calculated values for the magnetic field, which is used to determine the momenta of muons during the detector operation. For such measurements the field values should be known to a few percent. For this purpose, 405-turn flux-loops have been installed around selected regions of the CMS yoke plates to permit the measurement of magnetic flux density induced in the steel when the field in the solenoid is changed [4].

A few dozens of 3-D Hall sensors will be installed to measure and monitor the magnetic field near the surfaces of steel plates in small air gaps between the barrel wheels, in large air gaps between barrel and the first end-cap disks, and between the end-cap discs. The field values measured with these Hall probes will be used to estimate the remanent field in the yoke steel after the solenoid discharges and also to calibrate the calculations made with the CMS magnetic field model. The model calculations will be used then everywhere in the detector volume except of the region inside the solenoid bore where the field will be measured with 0.1% accuracy using a field mapping device.

To investigate the possibility of measuring the average magnetic flux density in the selected cross sections of the CMS yoke plates with an accuracy of a few percent, a special R&D program was performed with several samples made of the CMS yoke steel. A sample, with a test flux-loop around its periphery and Hall sensors on its surface, was placed into a slowly increasing or decreasing external magnetic flux produced by a laboratory dipole electromagnet (called "test-magnet" in what follows) connected to a computer-controlled power supply.

The value of the magnetic flux in the sample is determined from a combination of the flux-loop and Hall sensors mounted on the sample as described in this paper.

Manuscript received October 20, 2003; revised March 26, 2004. This work was supported in part by the U.S. Department of Energy under Contract DE-AC02-76CHO3000.

V. I. Klyukhin is with the Skobeltsyn Institute of Nuclear Physics, Moscow State University, Moscow, RU-119992, Russia, and with CERN, Geneva CH-1211, Switzerland (e-mail: Vyacheslav.Klyukhin@cern.ch).

D. Campi, B. Curé, A. Gaddi, H. Gerwig, J. P. Grillet, and A. Hervé, are with CERN, Geneva CH-1211, Switzerland.

R. Loveless is with the Department of Physics, University of Wisconsin, Madison, WI 53706 USA.

R. P. Smith is with the Fermi National Accelerator Laboratory, Batavia, IL 60510-0500 USA (e-mail: rpsmith@fnal.gov).

Digital Object Identifier 10.1109/TNS.2004.834722

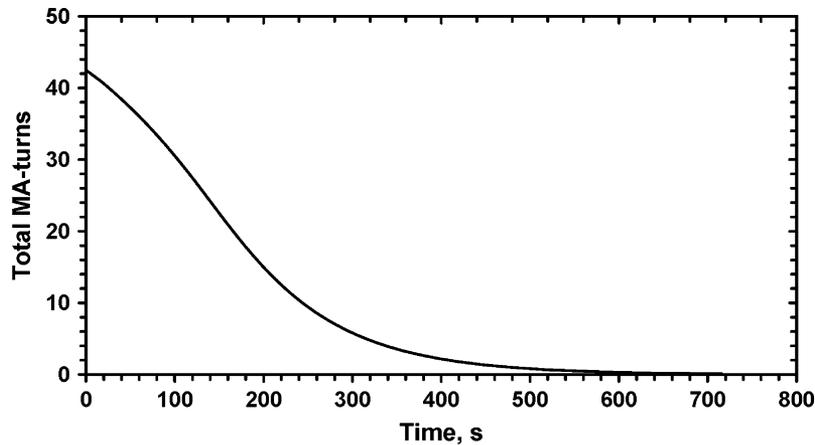

Fig. 1. Calculated CMS coil fast discharge.

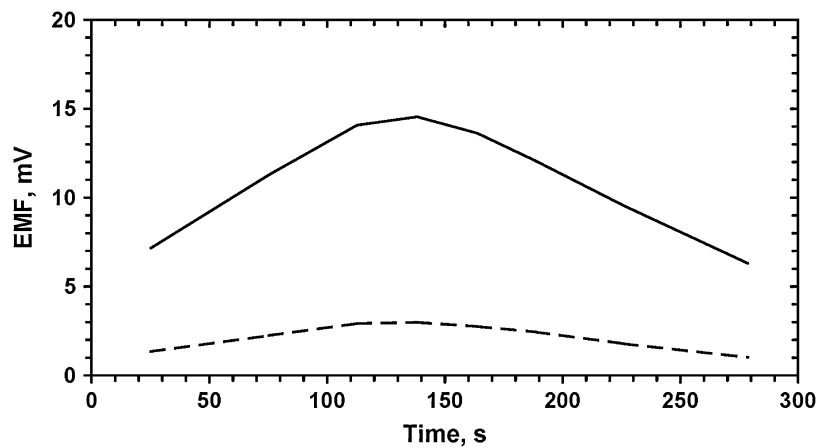

Fig. 2. Calculated minimum (dashed) and maximum (solid) EMF per one-turn flux-loop on the CMS barrel sectors.

## II. Modeling the CMS Fast Discharge

The normal charge or discharge rate of the CMS superconducting coil (5 hours) is too slow to induce usefully significant voltages in practical flux-loops. However, the rapid discharge of the solenoid (190 s time constant) made possible by the protection system which is provided to protect the magnet in the event of major faults, will induce significant voltages in the flux-loops installed in the CMS steel yoke. This protection system will be tested during the commissioning of the CMS magnet and this test provides an opportunity to measure the magnetic flux in the selected CMS steel yoke cross sections by offline software integration of the voltages measured in the flux-loops.

The rapid discharge of the CMS coil into the protection resistor provided for this purpose will cause quenching of the superconducting solenoid due to eddy current heating of the aluminum alloy coil support bobbin on the outside of the coil [5]. Thus, the discharge (Fig. 1) departs modestly from a simple L/R decay of an inductor into a fixed resistance.

This discharge results in flux changes in the various elements of the steel yoke of the CMS magnet. At 9 discrete times (0, 50, 100, 125, 151, 176, 200, 251, and 306 s) during the above discharge the fields were calculated in the detector. In the plates of the barrel and end-cap ferromagnetic parts of the yoke, the resulting flux density values were integrated over the areas enclosed by the flux-loops. These areas vary from 0.34 to 1.86 m$^2$ on the barrel sectors and from 0.59 to 1.42 m$^2$ on 18° segments of the end-cap disks. From the total flux $\Phi$ enclosed by each flux-loop the average voltages $V = \Delta\Phi/\Delta t$ induced in the loops by the flux changes between time intervals were calculated as shown in Figs. 2 and 3. The presence of eddy currents in the steel, and their effects on the induced voltages in the flux loops, has been ignored in the calculations of Figs. 2 and 3.

Because eddy currents are driven by the same flux changes that generate the voltages in the flux loops, their most significant contribution to magnetic flux changes occurs at times far from the beginning and end of the fast discharge. The gaps between steel plates in the CMS barrel wheels tend to decrease eddy currents between the plates.

Thus, the voltages induced in the flux-loops can be integrated with good precision to obtain the changes in average magnetic flux density in the selected cross sections of the CMS yoke.

## III. Details of the CMS Flux-Loops and Definition of the Test Program

Each CMS flux-loop is made of 9 turns of 45-conductor ribbon cable wound into a shallow groove machined into the peripheral surface of the steel plate that will be sampled. By connecting the two ends of the ribbon cable so that the individual conductors in the ribbon are offset by one conductor, a 405-turn flux loop is formed that encircles the selected part of the yoke. As can be seen in the above-calculated EMF estimates shown in Figs. 2 and 3, voltages peaking to several volts will be induced in the CMS flux-loops during the fast discharge of the solenoid.

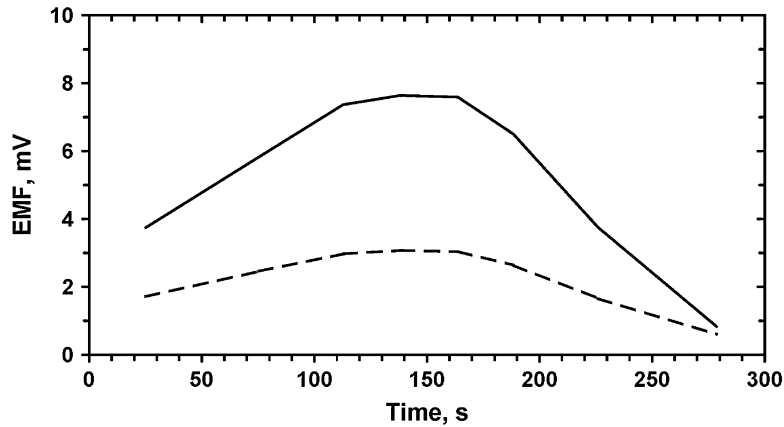

Fig. 3. Calculated minimum (dashed) and maximum (solid) EMF per one-turn flux-loop on 18° segments of the CMS end-cap disks.

To verify if these voltages can be measured online and integrated offline over the entire discharge to provide a measurement of the flux density change in the steel to an accuracy of a few percent, several sample discs 127 mm in diameter and 12.7 or 38.1 mm thick were made of the CMS yoke steel. Each sample disk was exposed to the magnetic flux produced by a test magnet discharged from a maximum current of 320 A with a current shape similar to the shape of the CMS current generated by the solenoid fast discharge.

The induced voltages were measured in a test flux-loop mounted on the sample disk. To provide an equivalent change in the magnetic flux, the number of turns of the test flux-loop was chosen larger than the number of turns in the CMS flux-loops, and the duration of the test magnet discharge was shorter than the CMS fast discharge time.

## IV. EXPERIMENTAL APPARATUS

### A. Voltage Measuring System

To measure voltages on the flux-loop a precision voltage sampling data acquisition system with fast ADC readout is used. This approach avoids the requirement for highly stable integrators that would otherwise be necessary to integrate the expected signals over the very long times of the CMS discharge.

A commercial multifunction analogue data acquisition circuit (National Instrument's DAQ Card 6012, fully encapsulated on a standard PCMCIA Type II card) provided 8 double-ended channels with 16-bit resolution. It was operated at 20 Hz, with absolute accuracy the order of 0.005%, using National Instrument's Labview software. The input impedance of the ADC is the order of 10 G$\Omega$ in parallel with 100 pF.

### B. Test Flux-Loop

The 994-turn test flux-loop was 140.6 mm in average diameter. It was wound on a nonmetallic flux-loop bobbin and connected to the sampling circuitry in differential mode to reject common-mode noise. The high-input impedance of the ADC system voltage insures that negligible current flows in the test flux-loop at any time.

The differential inputs of the ADC system were referenced to ground through 100 k$\Omega$ resistors. When the flux-loop system was tested in the laboratory environment, the noise signals were below 1 mV. The flux-loop was clearly sensitive to the nearby motion of a small permanent magnet at the 1–2 mV level.

### C. External Magnetic Flux

The test flux-loop was mounted between the iron pole tips of a test magnet (GMW Model 3474, energized with Danphysik Model 8530 power supply equipped with GPIB control interface), and the test magnet was charged and discharged at a number of different rates under control by the same software used to sample the voltage on the test flux-loop. The diameter of the test flux-loop was chosen so the entire flux-loop fit within the flat portion of the pole tips of the test magnet.

Numerous sets of measurements were performed with the pole tip gap of the test magnet set to 12.7 mm and 44.45 mm.

Three cases were investigated with the flux-loop inserted into the 12.7 mm gap of the test magnet: 1) the flux-loop was free of any inserted material; 2) an aluminum disk 12.7 mm thick and 127 mm in diameter was inserted into the flux-loop; and 3) a like-sized disk made from construction steel was inserted into the flux-loop.

Additional studies were conducted wherein sample disks of 38.1 mm thickness made of different steel used in the CMS yoke were inserted in a 44.45 mm gap of the test magnet. In these studies 3.175 mm air gaps between the samples and the pole tips of the test magnet existed on both sides of the disk, and Hall sensors were mounted on both sides of the disks at the centers of each side in the air gaps. The Hall sensors measured the axial magnetic flux density in the steel when the test magnet was fully energized, and the remanent field in the steel at the end of the discharge.

## V. MODELING THE TEST MAGNET

A TOSCA [6] model for the test magnet was prepared to guide the interpretation of the data obtained from the test flux-loop. The B-H data for the test magnet pole tips and yoke were taken from the measurements of the CMS yoke steel. This contributes only a small uncertainty into the calculations because at full current the steel sample disks and the test magnet pole tips were fully saturated.

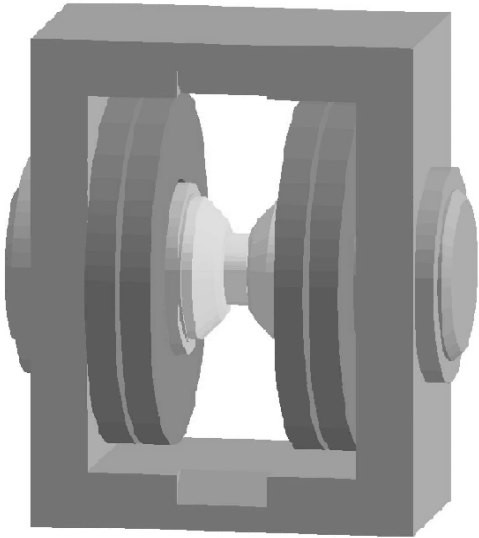

Fig. 4. TOSCA model for the test magnet with a 127 mm diameter and 38.1 mm thick steel sample disk inserted between the pole tips.

The TOSCA model shown in Fig. 4 predicts closely the flux density (2.65 T) in the 12.7 mm free air gap between the pole tips versus that measured by Hall probes (2.63 T) positioned on the surface of the pole tips when the test magnet was energized to full excitation. With the 12.7 mm thick steel sample disk inserted in the gap the model predicts a field of 3.07 T in the center of the disk. For both the air-gap and iron-filled gap the predicted magnetic flux density rises with increasing radial distance from the center of the pole tips. The magnetic flux density drops quickly outside the steel sample disk radii.

For the case with the 44.45 mm gap filled with the 38.1 mm steel disk spaced from the pole tips by two equal air gaps, the model predicts an axial field of 3.0 T in the center of the flat side of the disk when the test magnet is fully energized, whereas the Hall probes measure $2.9397 \pm 0.0002$ T. Based on this a correction factor of 0.9799 is applied to other calculated values when these are compared with the measured field values.

## VI. DATA FROM THE TEST FLUX-LOOP

### A. Test With a Small Gap Between Pole Tips

First, the test flux-loop was inserted in the gap of 12.7 mm between the pole tips and the test magnet charged to full current of 320 A at a charge rate of 2.5 A/s. After a pause, the current was decreased at the same rate to zero.

The voltage on the test flux-loop was sampled at 50 ms intervals (20 Hz sampling rate), and integrated offline by multiplying the average voltage in each time interval by the length of time interval. The time-integral of the voltage during charge-up is just the total flux change in the flux-loop, 40.91 Wb. During discharge the integral was 40.02 Wb. The measured flux is renormalized to flux density (e.g., 40.91 Wb corresponds to 2.65 T) using the area of the test flux-loop and the number of turns in the flux-loop. The measured flux changes from charging and discharging agree within 2%. The difference of these fluxes is attributed to the hysteresis of the test magnet poles, demonstrating that the measurement of the remanent field in the air gap is necessary to reconstruct precisely the magnetic flux density at full excitation of the test magnet.

Upon discharge, the voltage in the test flux-loop did not return to zero when the test magnet current reached zero but did so only after an additional 3–4 s. The voltage integral increased measurably during this extra time. The pole tips of the test magnet apparently demagnetize themselves shortly after the current reaches zero in response to the (rather small) demagnetizing fields remaining in the low carbon steel yoke of the test magnet.

Then, an aluminum disk 12.7 mm thick was placed in the flux-loop and the assembly inserted between the pole tips. The behavior of the voltage induced in the test flux-loop did not change showing that substantial eddy currents in the metal sample disks are excluded. This was also verified by comparing the induced voltages when the sample disk of construction steel was used versus those when the same sample disk was later divided into insulated quarters and inserted in the flux-loop.

### B. Test With Thick Steel Sample Disks in the Test Flux-Loop

The main studies were performed with two 38.1 mm thick sample disks made from the same steel as the majority of the CMS barrel yoke. Each was inserted into the test flux-loop and spaced from the test magnet pole tips by air gaps of 3.175 mm. In these studies the charge-up of the test magnet was always at the rate of 2.5 A/s, as shown in Fig. 5. Fast discharges were studied where the overall discharge times were 32, 64, 128, 256, and 512 s, following shapes similar to the CMS fast discharge. Fig. 6 shows the induced voltage and integrated magnetic flux density for the discharge time of 32 s.

As shown in Fig. 5, the voltages recorded on the test flux-loop show a $\sim 2.3$ V peak early in the charge-up, corresponding to the rapid increase of magnetization of the steel disk with increasing test magnet current. At the end of the charge-up the magnetic flux density at the Hall probes placed in the center of the flat sides of the sample disk was measured.

Before the charge-up and at the end of the discharge the Hall probes measured the remanent fields in the air gaps, $B_{r\,ch}$ and $B_{r\,dis}$, respectively. It was observed that these remanent fields increased for longer discharge times. It was 37 mT for 32 s discharges and increased to 59 mT for 512 s discharges. The eddy currents in the test magnet poles cause this effect, and it results in a long tail of the induced voltage after the current of the test magnet is set to zero. For all the discharges this tail was measured during 70 s after $t = 0$ (after 32–512 s from the beginning of the discharge), where $t = 0$ represents the time when the current was requested by software control to become zero. In the case of the shortest discharge of 32 s this tail contributes 1.8% to the integrated voltage. For the 512 s discharge this contribution was 0.008%. A detailed examination of the induced voltage tail near the end of the 32 s discharge is shown in Fig. 7.

The test magnet charge-up and discharge occur as a series of small discrete steps in current visible in Figs. 5, 6, and 7.

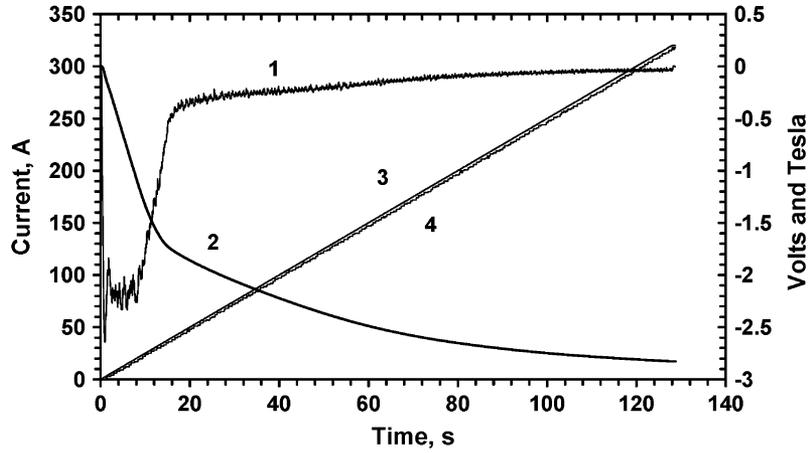

Fig. 5. Induced voltage (curve 1) and the integrated magnetic flux density (curve 2) versus time as the test magnet charges to 320 A at 2.5 A/s. Curve 3 shows the current request from the control software. Curve 4 corresponds to the measured current read-back.

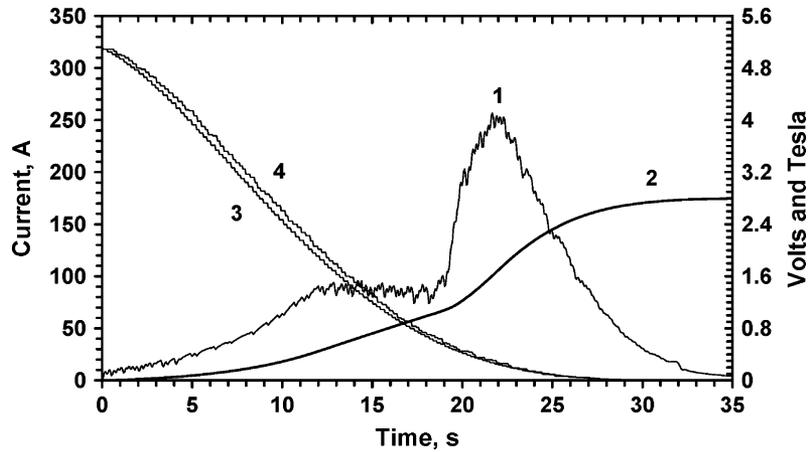

Fig. 6. Induced voltage (curve 1) and the integrated magnetic flux density (curve 2) when the test magnet ramped from 320 A to zero current during 32 s. Curve 3 shows the current request from the control software. Curve 4 corresponds to the measured current read-back.

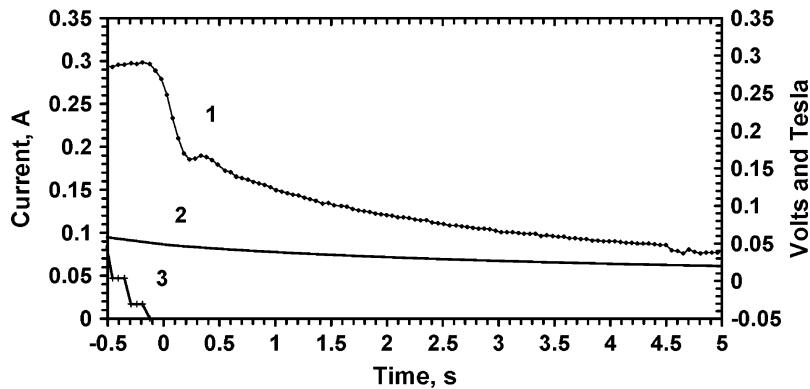

Fig. 7. Induced voltage (curve 1) during the first 5 s after $t = 0$ (after 32 s from the beginning of the discharge), where $t = 0$ represents the time when the requested current became zero (curve 3). Curve 2 (the magnetic flux density integral) has been "inverted" by subtracting each instantaneous value of the integral from the final value of the integral measured 70 s after the test magnet current was set to zero ($t = 0$ in the plot).

The voltage induced in the test flux-loop, although smoothed by the inductance of the test magnet, shows this influence as displayed in Fig. 8. The precision of the current measuring hardware is approximately 1 A. Actual resistive discharges of the CMS magnet will be completely smooth and this influence will disappear.

### C. Results of Measuring the Magnetic Induction in Steel

In 11 charge/discharge cycles of varying discharge times the sums $B_{r\,ch} + B_{i\,ch}$, and $B_{i\,dis} + B_{r\,dis}$ were investigated. In these sums $B_{i\,ch}$ is the magnetic flux density obtained from the magnetic flux integrated by the test flux-loop during the charge-up of the test magnet, and $B_{r\,ch}$ is the remanent field measured by the Hall sensors before the charge-up began. The subscript "dis" denotes the same quantities measured during discharges of the test magnet (including 70 s after the current was ramped to zero) with the Hall sensor value recorded after the discharge. Averaging the results from the 11 different cycles gives the values $\langle B_{r\,ch} + B_{i\,ch}\rangle = 2.8633 \pm 0.0018$ T for charging and $\langle B_{i\,dis} + B_{r\,dis}\rangle = 2.8583 \pm 0.0028$ T for discharging. The results agree within 0.2%.

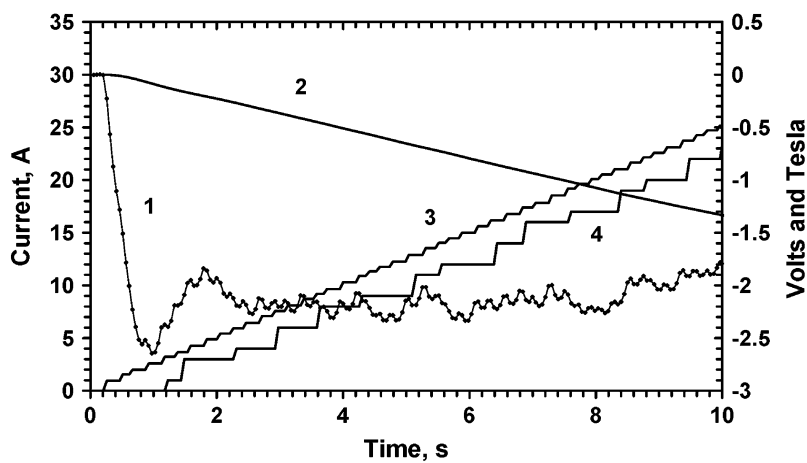

Fig. 8. Induced voltage (curve 1) and the integrated magnetic flux density (curve 2) versus time during the first 10 s of the test magnet charge-up. Curve 3 describes the current request from the control software. Curve 4 corresponds to the measured current read-back.

Taking the TOSCA calculations for the flux-loop and scaling by the correction factor (described previously), a calculated magnetic flux density of 2.8726 T is obtained. This agrees with $\langle B_{r\,ch} + B_{i\,ch}\rangle$ within 0.3% and with $\langle B_{i\,dis} + B_{r\,dis}\rangle$ within 0.5%.

The good agreement of these measurements with TOSCA calculations as well as the equality of $\langle B_{r\,ch} + B_{i\,ch}\rangle$ and $\langle B_{i\,dis} + B_{r\,dis}\rangle$ indicate that the precision of the measurements is within 0.5%.

## VII. CONCLUSION

The experimental program described herein indicates that the magnetic flux density in a steel object magnetized by an external source can be measured with good precision using a combination of the flux-loops and Hall probes.

Using the data from the flux loops and the Hall probes mounted on the selected regions of the CMS yoke, the magnetization of the steel can be measured with these techniques to a precision comfortably within a few percent.